\documentclass[iop]{emulateapj}

\usepackage{amsmath, amsthm}
\usepackage{booktabs}
\usepackage{xcolor}

\begin{document}

\shortauthors{S. Pellegrini et al.}
\shorttitle{X-ray haloes and AGN feedback}

\title{AGN feedback and the origin and fate of the hot gas in early-type galaxies}

\author{Silvia Pellegrini}
\affiliation{Department of Physics and Astronomy, University of Bologna, via Gobetti 93/2, 40129 Bologna, Italy}
\author{Luca Ciotti}
\affiliation{Department of Physics and Astronomy, University of Bologna, via Gobetti 93/2, 40129 Bologna, Italy}
\author{Andrea Negri}
\affiliation{Instituto de Astrof{\'i}sica de Canarias, calle V{\'i}a L{\'a}ctea, E-38205 La Laguna, Tenerife, Spain}
\author {Jeremiah P. Ostriker}
\affiliation{Department of Astronomy, Columbia University, 550 West 120th St, New York, NY 10027, USA}

%%% Definitions:
\newcommand*{\Lx}{L_{\rm X}}
\newcommand*{\Tx}{T_{\rm X}}
\newcommand*{\Mst}{M_{\star}}
\newcommand*{\Mt}{M_{\rm t}}
\newcommand*{\MR}{$M_{{\rm t,5}R_{\rm e}}$}

\begin{abstract}
A recent determination of the relationships between the X-ray
luminosity of the ISM ($\Lx$) and the stellar and total mass, for a
sample of nearby early-type galaxies (ETGs), is used to investigate the origin of
the hot gas,  via a comparison with the results of hydrodynamical simulations
of the ISM evolution for a large set of isolated ETGs. After the epoch of
major galaxy formation (after $z\simeq 2$),  the ISM is replenished
by stellar mass losses and SN ejecta, at the rate predicted by
stellar evolution, and is depleted by star formation; it is heated
by the thermalization of stellar motions, SNe explosions and the mechanical (from winds) and radiative AGN
feedback. The models agree well with the observed relations,
even for the largely different $\Lx$ values at the same mass, thanks to the 
sensitivity of the gas flow to many galaxy properties; this holds for models
including AGN feedback, and those without. Therefore, the mass input from the 
stellar population is able to account for  a major part of the observed $\Lx$; and 
AGN feedback,  while very important to maintain massive ETGs in a time-averaged quasi-steady state, 
keeping low star formation and the
 black hole mass, does not dramatically alter the gas content originating in stellar recycled material.
These conclusions are based on theoretical predictions
for the stellar population contributions in mass and energy, and on a self-consistent 
modeling of AGN feedback. 
\end{abstract}


\section{Introduction} 

In the recent years our knowledge of the dynamical structure of
early-type galaxies (hereafter ETGs), and of the properties of their
hot interstellar medium, has improved considerably.  On the one hand,
estimates of the luminous and dark mass distributions have become
possible out to large radii, thanks to integral field spectroscopy and
the use of mass tracers extending to many effective radii,
as globular clusters and planetary nebulae (e.g., Deason et al. 2012, Norris et
al. 2012, Morganti et al. 2013, Napolitano et al. 2014, Cappellari et
al. 2015, Alabi et al. 2016, 2017, Foster et al. 2016, Poci et
al. 2017; see also Cappellari 2016). On the other hand, a large number
of ETGs have had a significant exposure with the
X-ray satellite $Chandra$, and the resulting data have been analyzed
with uniform procedures by various groups,
obtaining properties for the hot gas more accurate ever; in particular,
for the first time the contribution of stellar sources to the total X-ray
emission could be efficiently removed [e.g., Boroson et al. 2011; Kim \&
Fabbiano 2013 (hereafter KF13), 2015; Su et al. 2015; Goulding et
al. 2016].

The improved knowledge of the dynamical structure of ETGs and of their
hot gas properties has been used to investigate the link between the
total mass (stars+dark matter, $\Mt$) and the hot gas
luminosity ($\Lx$) and temperature. In fact, it is expected
that the gas content and temperature depend primarily on $\Mt$, being
related respectively to the gas binding energy and virial temperature
(e.g., Ciotti et al. 1991, David et al. 1991, Pellegrini 2011, Posacki
et al. 2013, KF13). It was thus shown that indeed the total mass is the
primary factor in retaining the hot gas.
For a sample of 14 ETGs, the relation between $\Lx$ and the total mass
within $5R_{\rm e}$ (hereafter \MR), where $R_{\rm e}$ is the optical
effective radius, turned out to be tighter than that between $\Lx$ and
the total optical luminosity, especially for gas rich ETGs (KF13).
For ETGs with $\Lx<$few$\times 10^{40}$ erg
s$^{-1}$, the scatter in the relation suggests the importance of
secondary factors, as galaxy rotation, galaxy shape, star formation
history, environment, AGN feedback.  Forbes et al. (2017; hereafter F17) recently revisited the
correlation between $\Lx$ of ETGs and various masses: the
galaxy stellar mass ($\Mst$), \MR, and $\Mt$. They used homogeneous
mass measurements for 29 galaxies from the SLUGGS survey, that
provides kinematic information for stars (to a few $R_{\rm e}$) and
globular clusters (out to $\simeq 10R_{\rm e}$) in two dimensions (Alabi
et al. 2017). Consistently with the earlier results of KF13, they
found a strong linear relationship between $\Lx$ and \MR, confirming
that the total mass is the primary factor in driving the hot gas
content. F17 also compared their relationships 
with the predictions of SPH simulations of galaxy formation and subsequent evolution 
within the $\Lambda$CDM paradigm, where a major source for the ISM is provided by
infall within dark matter haloes of pristine gas, that is shock-heated to X-ray emitting temperatures.
In particular, they found agreement with the results
of Choi et al. (2015), for the set of simulations
that include mechanical and radiation heating from AGN (while models without feedback, or with 
thermal feedback, show systematically higher X-ray luminosity than observed, as already noted by 
Choi et al. 2015).

In another approach that makes use of hydrodynamical simulations,
but of the 'grid' type, the hot gas evolution is
followed after the epoch of galaxy formation (Ciotti et al. 1991; Negri
et al. 2014, hereafter N14; Ciotti et al. 2017, hereafter C17), starting with
galaxies almost empty of ISM, due to 
the combined effects of the SNe explosions and of
the AGN that are believed to clear the galaxies from the residual ISM, and 'quench'
star formation (e.g., Silk \& Rees 1998; Di Matteo et al. 2005, 2012; Debuhr
et al. 2012, Dubois et al. 2013; Vogelsberger et al. 2013; Khandai et
al. 2015; Schaye et al. 2015; Bieri et al. 2017; DeGraf et al. 2017; Barai et al. 2017). 
The ISM is then replenished by the collective input provided by the stellar population
during its normal ageing, via stellar mass losses and supernovae
ejecta.  This source of mass for the ISM is accurately predicted by the stellar evolution
theory, and it is not minor (of the order of $\approx 10$\% of $\Mst$;
e.g., C17).  The resulting ISM evolution has been
investigated for a large set of representative galaxy models,
varying their $\Mt$, their shape, internal kinematics, the presence of
AGN feedback and star formation [N14, Negri et al. 2015 (hereafter N15); C17].
Here we show how this modeling, at the present epoch, can account well
for the relations presented by F17, with implications for the origin of the hot gas observed today, and
the effect of AGN feedback.

The paper is organized as follows: in Sect. 2 we briefly describe the
numerical simulations (main code properties, model galaxies, mass and
energy input from the stellar population and the AGN); in Sect. 3 we
present the resulting relations between the hot gas properties and
the mass of the models on various scales, and compare them with those obtained
by F17 for observed ETGs; in Sect. 4 we discuss the results and present the
conclusions.

\section{The models: dynamical structure, stellar population inputs, AGN feedback}

We very briefly describe here the main features of the modeling and of the
numerical implementation adopted to describe the evolution of the hot
ISM in ETGs in our recent works (see N14, N15, C17 for more details on the specific 
simulations used here,  and Ciotti \& Ostriker 2012
for a detailed description of the realization of all input sources and of AGN feedback).

High resolution 2D hydrodynamical simulations, performed with the ZEUS
MP2 code, have been run for a large set of underlying axisymmetric two-component
(star+dark matter) galaxy models, with a central massive black hole (MBH), of shapes
ranging from E0 to E7.  The Jeans equations provide the internal
stellar kinematics, ranging from that of the isotropic rotator to the
fully anisotropic case, on which the stellar kinematical heating is
then based (N14). The stellar density profile follows a deprojected
(ellipsoidal) Sersic law, and the main observables (galaxy luminosity
$L$, $R_{\rm e}$, central stellar velocity dispersion) are related to
lie on the main scaling laws. The spherical dark matter halo has a
radial profile predicted by cosmological simulations (Navarro et
al. 1997), and is normalized to account for a dark-to-stellar mass
ratio of $M_h/\Mst\simeq 20$ (e.g., Behroozi et al. 2013), and for a dark mass
within $R_{\rm e}$ lower than the stellar mass, as measured in the
local universe (e.g., Cappellari et al. 2015).  With respect to ETGs of the local universe, the galaxy models built in this way
 are the most realistic ones adopted so far in hydrodynamical simulations (Posacki et al. 2013, N14);  they are
kept fixed during the time span of the simulations.
We do not account for the possibility of galaxy merging;
this is a less severe approximation than might seem, though, because the
stellar dynamical time and the sound crossing time are comparable.
Thus, the most conspicuous effects of the merging on the stellar structure and gas dynamics 
fade away within $\approx 1-2$ Gyr, a fiducial virialization time.

The hot gas originates from stellar mass losses and SN ejecta, and
evolves under the action of gravity, of cooling, and of various energy
sources: SNe explosions, thermalization of the
random and ordered kinetic energy of stellar motions, and accretion
onto the central MBHs, when present.  The
mass and energy input from the ageing stellar population are secularly
declining, according to the prescriptions of stellar evolution theory
for the 'normal' stellar mass losses, and, for the Type Ia supernova
rate, to the predictions of progenitors evolution and to survey
observations extending to medium redshift (e.g., Pellegrini
2012). Note that this modeling is still valid even in
presence of gas-poor merging, because the added galaxy will contribute with its own
stellar population inputs.
Star formation is also included, via a simple scheme based
on physical arguments shown to reproduce well the Kennicutt-Schmidt
relation (N15); cold gas is removed from the numerical grid due to
star formation, and type II supernovae produced by the newly born
stellar population are also considered, as sources of mass and energy
 [see Ciotti \& Ostriker (2012) for a thorough description of the mass
and energy inputs from the old, secularly evolving stellar population, and 
the newly born one, and of their realization in the code].

C17 followed the evolution of hot gas flows including an accurate and
physically self-consistent implementation of AGN feedback, both
radiative and mechanical, the latter due to AGN winds (as observed
for BAL AGNs and high-redshift quasars, e.g. Tombesi 2016, Zakamska et
al. 2016). The adopted grid modeling allows for a high resolution of the
hydrodynamical fields over the whole extent of the galaxy; with
a logarithmic grid spacing, the central resolution is 5 pc,
that is inside the fiducial accretion radius (e.g., Pellegrini 2010,
Ciotti \& Pellegrini 2017), and the outermost radius is 250 kpc.
 Moreover, the heating of the ISM by AGN feedback results from
the mass accretion rate on the MBH computed with a high central resolution,
a self-consistent treatment 
of the mass, energy and momentum balance of the inflowing and outflowing material at the first
radial gridpoint (Ostriker et al. 2010), and radiative and mechanical
efficiencies, including their variation with the mass accretion rate, in agreement with 
current observational and theoretical findings.
Sub-grid physics (i.e., accretion on a free-fall time) applies just to the innermost 
gridpoint,  inside the accretion radius. In conclusion, we
do not resort to assumptions to determine the mass accretion rate
(as, e.g.,  a ``boost factor''; Booth \& Schaye 2009, Dubois et al. 2015), and we cannot adjust 
the ``strength'' of AGN feedback, since the heating of the ISM resulting from the accretion
process is self-determined.

\begin{figure*}
\vskip -6.5truecm
\hskip -0.6truecm
\includegraphics[width=1.06\linewidth, keepaspectratio]{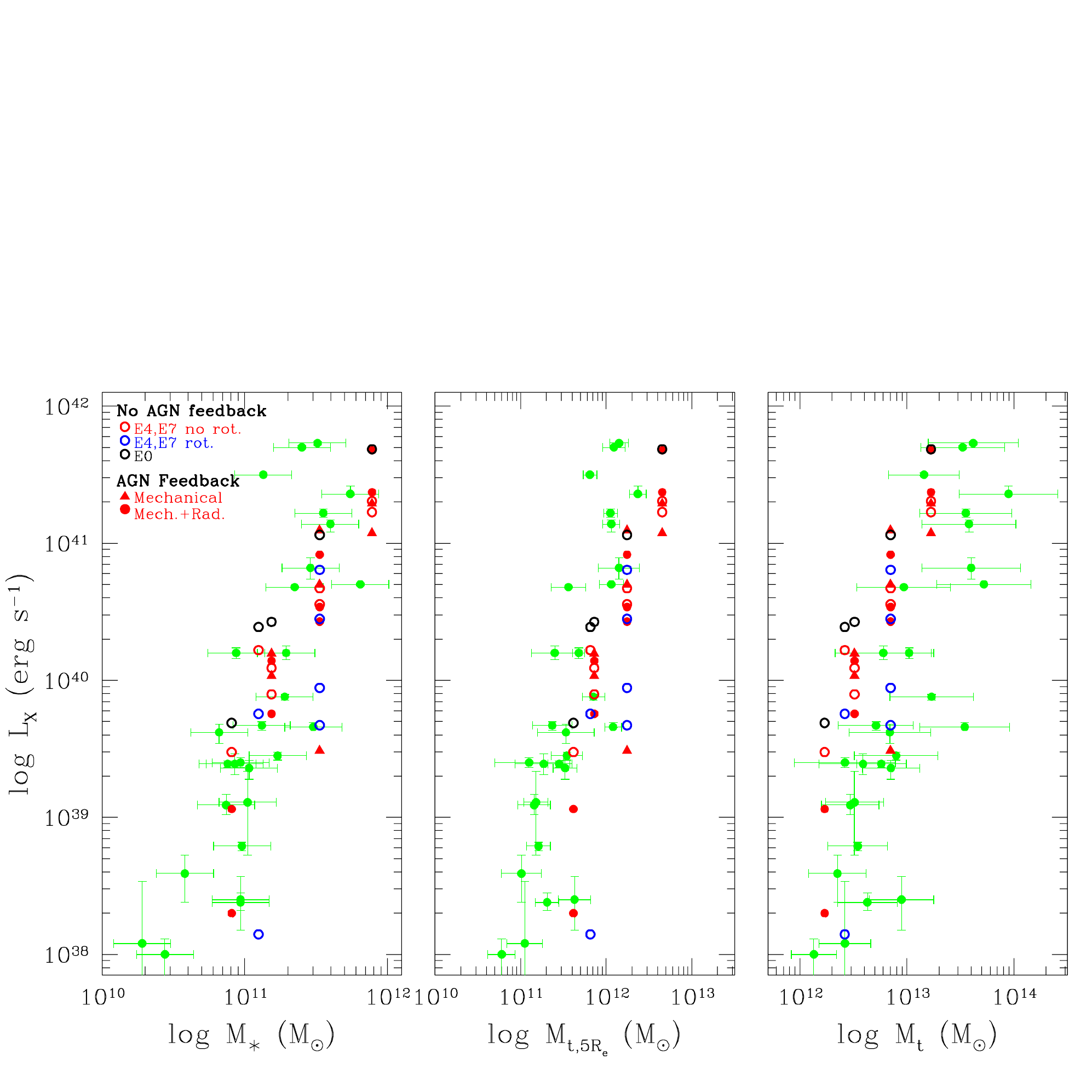}  % fatta da f1.m
\caption{Observed (green) and model X-ray luminosities versus the total stellar mass $\Mst$ (left), the total enclosed mass
within 5$R_{\rm e}$ \MR (middle), and the total virial mass $\Mt$ (right). For observed ETGs, X-ray luminosities and
masses are taken from F17, who derived stellar+dark masses from globular cluster kinematics (Sect. 3). 
The models have E0, E4 and E7 shapes; full symbols refer to non-rotating galaxies with AGN feedback (mechanical, or mechanical and radiative); open symbols
refer to simulations without AGN feedback, and in this case the E4 and E7 galaxies can be rotating (blue open circles) or not (red open 
circles). The internal kinematics of rotating models is that of isotropic rotators; for $\Mst=3.35\times 10^{11}$M$_{\odot}$ also two cases 
with a lower rotational level are plotted (these have the largest $\Lx$ among rotating models at this $\Mst$; N15).
See Sect. 2 for more details and references for the models.}
\label{f1}
\end{figure*}

Each galaxy evolves in isolation, and at the beginning of the
simulation is assumed to be $\simeq 2$ Gyr old, an age after which the
prescriptions for the stellar population inputs are accurate and
reliable (e.g., Pellegrini 2012); this epoch can be placed at around
$z\simeq 2$, since the bulk of the stellar population of ETGs is old
(e.g. Renzini 2006,  Prichard et al. 2017). At the beginning the galaxies are also assumed to
be devoid of gas, as a result of outflows powered by SNe and/or quasars
that ended the high star formation epoch (Sect. 1). The simulations follow the ISM evolution for the subsequent
11 Gyr, thus ending at a galaxy age of $\simeq 13$ Gyr.

For comparison with the X-ray observations and mass determinations in F17,
we consider here two sets of models, of shapes ranging from E0 to E4 and E7: the
first set includes models without AGN feedback (N14, N15, C17),  and the
internal kinematics for the E4 and E7 shapes may include rotation 
or not; the second set contains models with AGN feedback (mechanical or
radiative+mechanical), and the simulations have been performed for
non-rotating galaxies (C17). The role of AGN feedback in rotating galaxies is under investigation.
All details about the galaxy models and their X-ray properties are given in the references
above.

\section{The models in the $\Lx-$mass plots}
Figure 1 shows  the present-epoch values of $\Lx$ for the set of model galaxies described in Sect. 2,
versus the various masses
considered by F17: the total stellar mass $\Mst$, the total mass
within 5$R_{\rm e}$ (\MR), and the total galaxy mass $\Mt$. The X-ray
emission, in the energy band of 0.3--8 keV, comes from a sphere of
$5R_{\rm e}$ radius, an extraction region typically used in
observational works; it includes almost all of $\Lx$ in models and observations.

Figure 1 has been built for a direct comparison of the models with the
results of F17,  thus it mimics exactly Fig. 2 of F17. The green points with errorbars 
show the F17 sample of observed ETGs, all within a distance
of 27 Mpc. For these ETGs, $\Lx$ 
over 0.3--8 keV comes mostly from the analysis of $Chandra$ data
conducted by KF13 and Kim \& Fabbiano (2015), and has been extracted from galaxy radii
which vary from 2 to 5 $R_{\rm e}$.  The dynamical masses within
5$R_{\rm e}$ derive from the Globular Clusters kinematics of the
SLUGGS survey (Alabi et al. 2017); the total virial masses $\Mt$ are
calculated by extrapolating from \MR, assuming an NFW-like dark matter
halo.  While definitely minor within $R_{\rm e}$ (Cappellari et al. 2015),
the  dark matter fraction within
5$R_{\rm e}$ shows a range of values (from $\sim 0.1$ to 0.9),
for $\Mst\ga 10^{11}$M$_{\odot}$ (Alabi et al. 2017); between 5$R_{\rm e}$ and the
virial radius there is practically only dark mass. Thus, the dark mass resides 
mostly outside the region where the bulk of the stellar recycled material is produced.
Finally, the ETGs in the sample of F17 lie in a variety of environments, from the 
field to the group to the cluster.
Only two of them are central dominant galaxies in clusters (M87, not shown in Fig.~\ref{f1} for its large $\Lx$,
 and NGC1399); other two are optically luminous ETGs at the center of their groups (NGC5846 and NGC1407).

\subsection{General comparison with observations}
The left panel shows that $\Lx$ of the models reproduces well the
distribution of observed values, at all the $\Mst$ of the
simulations ($\Mst=0.81,\,1.25,\,1.54,\,3.35,\,7.80\times 10^{11}$M$_{\odot}$). 
From the lowest to the largest $\Mst$, the
following main features of the hot gas flow allow to account for the
observed $\Lx$.  Models of the lowest $\Mst$ ($\Mst\la 1.2\times
10^{11}$M$_{\odot}$) host a significant outflow/wind region, mostly
SN-driven (as already established previously; e.g., Pellegrini \&
Ciotti 1998, David et al. 2006, Pellegrini et al. 2007) and then their
$\Lx$ keeps low; indeed, a few of the models run at these lower $\Mst$
have $\Lx< 10^{38}$ erg s$^{-1}$, and then they do not appear in the
plot. For $\Mst=1.2\times 10^{11}$M$_{\odot}$ only models without AGN
feedback have been run, yet their $\Lx$ ranges 
from $1.4\times 10^{38}$ erg s$^{-1}$ to $2.7\times 10^{40}$
erg s$^{-1}$, and is able to cover the large $\Lx$ interval 
observed at this $\Mst$. For $\Mst\ge 1.5\times 10^{11}$M$_{\odot}$,
all models have $\Lx >$ a few$\times 10^{39}$ erg s$^{-1}$, due to the
increasing binding energy of the gas with increasing galaxy mass
(e.g., Pellegrini 2012, Posacki et al. 2013).  The most systematically
explored mass is $\Mst =3.3\times 10^{11}$M$_{\odot}$, where the
models examine the effects of all the major galaxy properties (shapes from
E0 to E7, with and without rotation); again, $\Lx$ spans from $3\times
10^{39}$ erg s$^{-1}$ to $2\times 10^{41}$ erg s$^{-1}$, and covers
most of the very large observed range.  At fixed galaxy mass, $\Lx$ is
lower for flatter morphologies, because the outflowing region is
larger (N14); $\Lx$ is also lower in rotating ETGs, since rotation
creates a cold disk and reduces $\Lx$ (N14, N15).  The largest
observed $\Lx$ values\footnote{The two largest observed $\Lx$ in
  Fig. 1 belong to NGC1399 and NGC5846, two ``central'' ETGs (Sect. 3),
  while the models refer to isolated ETGs. For central galaxies,  an intracluster/intragroup
  medium can produce a confinement effect that increases $\Lx$.}
of a few$\times 10^{41}$ erg s$^{-1}$ are reproduced by the models
with $\Mst=7.4\times 10^{11}$M$_{\odot}$.  Note how the range of
observed $\Lx$ tends to narrow at the largest $\Mst$, a property that
is shown also by the models. In the models, this is explained in part
by the exclusion of flat and rotating galaxies, because the most
massive ETGs tend to be round and non-rotating; and in part by the
progressive reduction of the size of an outflowing region as the
galaxy mass increases.  In summary, previous models available in the
literature, not tailored specifically for comparison with the F17
results, appear to naturally account for the observed trend
of $\Lx$ with $\Mst$ of F17.  The large variation in $\Lx$ accomplished by
the models at the same $\Mst$ lies in the sensitivity of the gas flow
to many factors characterizing ETGs, as galaxy shape, presence of
rotation, presence of star formation, and AGN feedback.

The agreement between the models and the results of F17 keeps good
also in the other two panels of Fig. 1, where the x-axis includes the
dark mass; the latter provides already a dominant contribution within
$5R_{\rm e}$ (middle panel). The models seem to have \MR
slightly larger than derived for the sample of observed ETGs, for the same $\Mst$ (middle
panel), while they tend to have $\Mt$ lower than evaluated from
extrapolation of the inner mass profile by F17 (right panel). Note
however that a mismatch of the size in Fig. 1 between the amount and
distribution of dark matter in the models and in the estimates of F17
is not expected to make $\Lx$ resulting from the simulations
unreliable; in fact, the region important for determining $\Lx$ is the
inner few $R_{\rm e}$, where the stellar mass dominates and the models
have been built with the correct fraction of dark matter (Sect. 2,
N14). Moreover, the total mass estimates of F17 would imply that the gas should be
slightly less bound than in the models, within $5R_{\rm e}$, but
more bound on the largest scale; thus, it is not straightforward to
predict whether and in what sense $\Lx$ of the models would change,
when inserting exactly the F17 estimates.

\subsection{What is the role of AGN feedback?}
Here we give a closer inspection of Fig. 1 at the $\Mst$ values for which
models with and without AGN feedback have been run (full and open
symbols respectively), to understand whether there is a difference in
their relative distribution.  At the lowest $\Mst=8.1\times
10^{10}$M$_{\odot}$, the models in general host large outflowing
regions, due to the SNe heating, even without AGN feedback; the effect
of the latter is that of making the loss of gas easier, and then of a
further decrease of $\Lx$. For larger $\Mst\ge 1.5\times
10^{11}$M$_{\odot}$, the range of $\Lx$ covered by the three types of
models (no AGN feedback, mechanical feedback, and radiative+mechanical
feedback) is similar, at each $\Mst$; there is no clear tendency of
one type to occupy preferentially lower or larger $\Lx$ values.  The only feature
common at all $\Mst$ is that spherical models without AGN feedback are
always found at the top of the distribution of $\Lx$ values.

The similarity of $\Lx$ shown by models with and without AGN feedback
is the result of two important properties of the secular gas flow
evolution. The first is that the outburst episodes triggered by
accretion on the MBH are very brief: during them $\Lx$ of the hot gas
becomes very high, mostly due to a central bright and hot region, but,
after few$\times 10^7$ yr, $\Lx$ quickly drops, and most of the time
keeps at normal, ``quiescent'' values (C17). For example, during the
past 3--5 Gyr, the fraction of time spent above an $\Lx$
representative of a large gas emission observed for local ETGs 
($\Lx = 3 \times 10^{41}$ erg s$^{-1}$; Fig. 1) is just
$\approx 1$\% (C17; Pellegrini et al. 2012).  The second feature
important to understand the similarity of $\Lx$ in models with and
without AGN feedback is the actual effect of the latter on the gas
flow on the galactic scale.  After the first $\simeq 2$ Gyr of life of
ETGs, during the epoch that is simulated, adding AGN feedback produces
an increase in the ejected mass from the galaxy of the
order of 20\%$-$40\%, going respectively from the largest to the smallest modeled ETGs;
AGN feedback is not capable of
producing a global/major outflow by itself. Recurrent feedback helps
to temporarily displace the gas from the center (out to a radius of
$\la 10$ kpc), thus temporarily reducing $\Lx$ even considerably, but
it does not clear the whole galaxy from the gas. 
The net result is a comparable X-ray emission in present day ETGs that have
evolved with or without AGN feedback.  

Even though it does not dramatically alter the overall hot gas
content, AGN feedback produces a big difference
in the amount of gas accreted to the center over the galaxies'
lifetime, with important astrophysical consequences: during the
galaxy evolution after $z\simeq 2$,  it maintains ETGs in a time-averaged
quasi-steady state, and keeps central star formation and the MBH mass
low, by suppressing recurrent ``cooling flows''.  For example, feedback is
crucial to obtain final MBH masses close to the values observed (C17): in
models without AGN feedback, the final MBH mass is far larger than
measured in the local universe [e.g., in the Magorrian et al. (1998)
relation]. In the feedback models, instead,  after most of
the MBH mass has grown by the end of the quasar epoch, as in the
  Soltan argument, just $\la 3$\% of the total injected mass from stars is accreted on
the MBH, during the simulations, and the average increase in the MBH mass  is
just of a factor of few.  In summary, in lower mass
ETGs ($\Mst\la 10^{11}$M$_{\odot}$) most of the injected stellar mass
loss is expelled in an outflow, driven mainly by SNIa's explosions (as
already known; e.g., Ciotti et al. 1991, Pellegrini \& Ciotti 1998);
in models of larger mass, roughly half of the total mass losses since
an age of $\simeq 2$ Gyr ends recycled into new stars\footnote{The AGN
  feedback has also a positive action for the formation of new stars, which is 
currently referred to as ``positive feedback''
  (Ciotti \& Ostriker 2007, C17).}, and the other half goes for a minor fraction into the MBH,
and for a major fraction is ejected from the galaxy, due to SN heating
and the further help of the AGN feedback action.

\section{Discussion and conclusions}

We have compared a recent determination of the relations
between the observed $\Lx$ and the stellar and total mass, for a sample of local ETGs (F17), with the 
$\Lx$ values at the present epoch of our hydrodynamical modeling of the hot gas evolution in ETGs.
The model galaxies are isolated, and devoid of ISM at the beginning, as expected after the 
'quenching' phase that ended the major star formation process. The ISM is then replenished in mass according to
the prescriptions of the stellar evolution theory. 
With respect to previous numerical investigations of the fate of the recycled stellar material in ETGs,
 this work has two objectives:
1) to examine the possibility for the stellar input to account for a major part of the 
observed $\Lx$, based on the comparison of the model results at the 
present epoch with the new, above-mentioned relations; and
2) to clarify the possibility to use the present-day $\Lx$ as diagnostic of the impact of AGN 
activity, within a scenario where the hot gas mostly originates from the stellar population.
To achieve these scopes, we considered a large set of galaxy models, built to have 
the present-day structural and kinematical properties observed for ETGs. We 
remark that the model predictions are based on high resolution grid-type 
simulations, on a self-consistent computation of the mass accretion rate and of the heating of the gas
due to AGN feedback, and on a contribution 
of stellar evolution reproducing the prescriptions of theory 
(see, e.g., the secular behavior of the mass losses from the 
ageing stellar population, and of the Type Ia SNe explosions; the inclusion of star 
formation from the cooling hot gas, with mass and energy return from 
the newly born stellar population, including Type II SNe).

We have found a good agreement between the $\Lx-\Mst$ relation of the models and that
of observed ETGs, over the whole $\Mst$ range. In particular, the
largely different observed $\Lx$ values at the same $\Mst$ are
reproduced by the sensitivity of the gas flow to many properties
characterizing ETGs, as galaxy shape, presence of rotation, presence
of star formation, and AGN feedback.  The models also show an overall
agreement with the observed $\Lx-$\MR relation, though \MR of the models is
slightly larger than derived by F17, for ETGs of the same $\Mst$. 
The agreement between models and observations holds
with and without radiative and
mechanical (from AGN winds) AGN feedback. These results imply that the mass input
from the stellar population is able to account for a large part of
the observed $\Lx$, and that  -- under the assumptions of this work -- AGN feedback 
does not have a significant role in determining the present epoch $\Lx$.
This is an important conclusion about the fate of recycled stellar material in 
presence of AGN feedback, since an outcome of the simulations could have been, for
example,  a major
degassing of the galaxies.
Such a different outcome would have implied the {\it necessity} to assume other sources of 
mass  (as gas accretion from outside).

F17 noticed a good agreement between their results and the predictions
of cosmological SPH simulations, where the hot gas mostly originates
from a cosmological reservoir from which
gas falls into the dark matter halos, is shock-heated, and is then
further significantly heated by mechanical feedback due to AGN winds
(Choi et al. 2015).  If this mechanical feedback is not included,
the simulations predict an $\Lx$ far larger than observed; thus, it is concluded that the AGN plays an
important role in determining $\Lx$ (in addition to $\Mt$).
This conclusion is not in contradiction with the
results of the present work, that starts
{\it after} $z \simeq 2$, and indicates that the AGN has not shaped dramatically the value
of $\Lx$ observed today. In fact  the cosmological simulations of Choi et al. (2015; as of many 
others, see, e.g., Sect. 1), start from the galaxy formation epoch, and
obtain at early times AGN-driven powerful outflows that remove nearly
all the gas, and stop star formation. In this sense the role of the
AGN is certainly very important. Our simulations (N14, N15, C17)  start from these "initial
conditions" created at the epoch of quenching, and find that at later epochs the energy injected by the
AGN is not crucial in determining $\Lx$  of the recycled material
(while {\it fundamental} to stop the recurrent cooling flow and keep low the MBH mass).  The conclusions of the two 
approaches, one including and the other excluding the epoch of galaxy 
formation  and the presence of cosmological gas, then agree,  provided that
 AGN feedback in the first approach is stronger than found here. In fact, if
AGN-driven outflows are to clear the gas out of massive ETGs at early epochs
and/or prevent a substantial cosmological accretion, this
requires an efficiency of energy transfer from the AGN to the ISM 
much larger than found by C17 at low redshift, when AGN feedback is not even able to clear the galaxies
from the stellar mass losses. It could be that AGN feedback at the epoch
of galaxy formation was stronger,  perhaps due to a denser and more optically thick gas, and a
shallower galactic potential well. 

Finally we discuss further the different origin of the hot gas, 
in the stellar population or in the cosmological infall.
The input of mass considered in our simulation, 
from supernovae and stellar winds, including
their secular trend, and both for the ageing and the newly born stars,
is the minimum to be included for ETGs.  
The question then is how much infall  can be accomodated at low redshift
to still reproduce the observed $\Lx$.  If cosmological
infall is very important,  contributing to a mass within a few $R_{\rm e}$
much larger than that provided by stars, then the effects of AGN feedback should again be
stronger than found by C17, to obtain $\Lx$ of the values observed,
MBH masses of the correct size, and limit star formation (see also Eisenreich et al. 2017).  One
possibility could be to include the additional effect of a jet,  even though it is expected
to represent the bulk of the energetic output when
the Eddington-scaled mass accretion rate falls below a low threshold 
(of $\simeq 0.01$, Yuan \& Narayan
2014). Indeed, this scaled rate seems to be low at the present epoch; 
for example, in the models of C17 it keeps below
$\approx 10^{-3}$ for $\approx 75$\% of the time after $z\simeq 2$, and it has 
been found similarly low in observational studies (Pellegrini 2005, 2010, Gallo et al. 2010). 
Jets are invoked to moderate the central cooling of the ISM, and of the intracluster medium
(e.g., McNamara \& Nulsen 2007; Heckman \&
Best 2014). How exactly the jet heating of the ISM works is still not
very well understood; however, the present results suggest that, for
isolated ETGs, radiative plus mechanical (from AGN winds) feedback
already allows to obtain $\Lx$ as observed, MBH masses on the
Magorrian relation, and the correct low level of star formation (N15,
C17). Further work including the confining action exerted by an
intragroup/intracluster medium, or cosmological infall, would be
helpful to assess more thoroughly the origin of the hot gas, and the role of AGN feedback in ETGs.

\bigskip
\acknowledgements
S. P. is grateful to Dong-Woo Kim for helpful discussions.


\begin{thebibliography}{99}

\bibitem[Alabi(2016)]{} Alabi, A. B., Forbes, D. A., Romanowsky, A. J., et al. 2016, MNRAS, 460, 3838

\bibitem[Alabi(2017)]{} Alabi, A. B., Forbes, D. A., Romanowsky, A. J., et al. 2017, \mnras, 468, 3949

\bibitem[Barai(2017)]{} Barai, P., Gallerani, S., Pallottini, A., et al. 2017, \mnras, arXiv:1707.03014

\bibitem[Behroozi(2013)]{} Behroozi, P. S., Wechsler, R. H., Conroy
  C. 2013, \apj, 770, 57 

\bibitem[Bieri(2017)]{} Bieri, R., Dubois, Y., Rosdahl, J., et al. 2017,\mnras, 464, 1854

\bibitem[Booth(2009)]{} Booth C. M., Schaye J., 2009, MNRAS, 398, 53

\bibitem[Boroson(2011)]{} Boroson, B., Kim, D.W., Fabbiano, G., 2011, \apj, 729, 12 

\bibitem[Cappellari(2015)]{} Cappellari, M., Romanowsky, A. J., Brodie, Jean P., et al. 2015, \apj, 804, L21

\bibitem[Cappellari(2016)]{} Cappellari, M. 2016, ARA\&A 54, 597

\bibitem[Choi(2015)]{} Choi, E., Ostriker, J.P., Naab, T., Oser, L., Moster, B. P. 2015, \mnras, 449, 4105

\bibitem[Ciotti(1991)]{} Ciotti, L., D'Ercole, A., Pellegrini, S.,  Renzini, A. 1991, \apj, 376, 380

\bibitem[Ciotti(2007)]{} Ciotti, L., Ostriker, J.P. 2007, \apj, 665, 1038

\bibitem[Ciotti(2012)]{} Ciotti, L., Ostriker, J.P. 2012, in {\it Hot Interstellar Matter in Elliptical
Galaxies},  Kim D.-W., Pellegrini S., eds, Astrophysics and Space Science Library, Vol. 378. Springer-Verlag, Berlin, p. 8

\bibitem[Ciotti(2017)]{} Ciotti, L., Pellegrini, S.,  Negri, A., Ostriker, J.P. 2017, \apj, 835, 15 (C17)

\bibitem[Ciotti(2017b)]{} Ciotti, L., Pellegrini, S.,  2017, \apj, 848, 29

\bibitem[David(1991)]{} David, L. P., Forman, W., \& Jones, C. 1991, \apj, 369, 121

\bibitem[David(2006)]{} David, L. P., Jones, C., Forman, W., Vargas, I. M., Nulsen, P. 2006, \apj, 653, 207

\bibitem[Deason(XXX)]{} Deason, A. J., Belokurov, V., Evans, N. W., \& McCarthy, I. G. 2012, \apj, 748, 2

\bibitem[Debuhr(2012)]{} Debuhr, J., Quataert, E., Ma, C.-P. 2012, \mnras, 420, 2221

\bibitem[DeGraf(2017)]{} DeGraf, C., Dekel, A., Gabor, J., Bournaud, F. 2017, \mnras, 466, 1462

\bibitem[Dimatteo(2005)]{} Di Matteo, T., Springel, V., Hernquist, L. 2005, Nature 433, 604

%\bibitem[Dimatteo(2008)]{} Di Matteo, T., Colberg,  J., Springel, V., Hernquist, L., Sijacki, D. 2008, \apj, 676, 33

\bibitem[Dimatteo(2012)]{} Di Matteo, T., Khandai, N., DeGraf, C., et al. 2012, \apj, 745, L29

%\bibitem[Dubois(2012)]{} Dubois Y., Devriendt J., Slyz A., Teyssier R., 2012, MNRAS, 420, 2662

\bibitem[Dubois(2013)]{} Dubois, Y., Gavazzi, R., Peirani, S., Silk, J. 2013, \mnras, 433, 3297

\bibitem[Dubois(2015)]{} Dubois, Y., Volonteri, M., Silk, J., Devriendt, J., Slyz, A., Teyssier, R. 2015, \mnras, 452, 1502

\bibitem[Eisenreich(2017)]{} Eisenreich, M., Naab, T., Choi, E., Ostriker, J.P., Emsellem, E. 2017, \mnras, 468, 751

\bibitem[Forbes(2017)]{} Forbes, D. A., Alabi, A., Romanowsky, A. J., et al. 2017, \mnras, 464, L26 (F17)

\bibitem[Gallo(2010)]{} Gallo, E., Treu, T., Marshall, P. J., et al. 2010, \apj, 714, 25

\bibitem[Goulding(2016)]{} Goulding, A.D., Greene, J.E., Ma, C.-P., et al. 2016, \apj, 826, 167

\bibitem[Heckman(2014)]{} Heckman, T.M., Best, P.N. 2014, ARA\&A, 52, 589

\bibitem[Kandai(2015)]{} Khandai, N., Di Matteo, T., Croft, R., et al. 2015, \mnras, 450, 1349

\bibitem[Kim(2013)]{} Kim, D.-W., \& Fabbiano, G. 2013, ApJ, 776, 116 (KF13)

\bibitem[Kim(2015)]{} Kim, D.-W., Fabbiano, G., 2015, \apj, 812, 127 

\bibitem[Magorrian(1998)]{} Magorrian, J. et al., 1998, AJ, 115, 2285

\bibitem[McNamara(2007)]{} McNamara, B. R., Nulsen, P. E. J. 2007, ARA\&A 45, 117

\bibitem[Napol(rr)]{} Napolitano, N. R., Pota, V., Romanowsky, A. J., et al. 2014, MNRAS, 439, 659

\bibitem[Navarro(1997)]{} Navarro, J.F., Frenk, C.S., White, S.D.M. 1997,  \apj, 490, 493

\bibitem[Negri(2014)]{} Negri, A., Posacki, S., Pellegrini, S., Ciotti, L. 2014, \mnras, 445, 1351 (N14)

\bibitem[Negri(2015)]{} Negri, A., Pellegrini, S., Ciotti, L. 2015, \mnras, 451, 1212 (N15) 

\bibitem[Norris(ccc)]{} Norris, M. A., Gebhardt, K., Sharples, R. M., et al. 2012, \mnras, 421, 1485

\bibitem[Ostriker(2010)]{} Ostriker, J.P., Choi, E., Ciotti, L., Novak, G., Proga, D. 2010, \apj, 722, 642

\bibitem[Pellegrini(1998)]{} Pellegrini, S., Ciotti, L. 1998, A\&A 333, 433

\bibitem[Pellegrini(2005)]{} Pellegrini, S. 2005, \apj, 624, 155

\bibitem[Pellegrini(2007)]{} Pellegrini, S., Baldi, A., Kim, D.W., Fabbiano, G., et al. 2007, \apj,667, 731

\bibitem[Pellegrini(2010)]{} Pellegrini, S. 2010, \apj, 717, 640

\bibitem[Pellegrini(2011)]{} Pellegrini, S. 2011, \apj, 738, 57

\bibitem[Pellegrini(2012a)]{} Pellegrini, S. 2012, in {\it Hot Interstellar Matter in Elliptical
Galaxies},  Kim D.-W., Pellegrini S., eds, Astrophysics
and Space Science Library, Vol. 378. Springer-Verlag, Berlin, p. 21

\bibitem[Pellegrini(2012b)]{} Pellegrini, S., Ciotti, L., Ostriker, J.P. 2012, \apj, 744, 21

\bibitem[Poci(2017)]{} Poci, A., Cappellari, M., McDermid, R. M. 2017, \mnras, 467, 1397

\bibitem[Posacki(2013)]{} Posacki, S., Pellegrini, S., Ciotti, L. 2013, \mnras, 433, 2259

\bibitem[Schaye(2015)]{} Schaye, J., Crain, R.A.; Bower, R.G., et al. 2015, \mnras 446, 521

\bibitem[Silk(1998)]{} Silk, J., Rees, M.J., 1998, A\&A 331, L1

\bibitem[Su(2015)]{} Su, Y., Irwin, J.A., White, R.E. III, Cooper, M.C. 2015, \apj, 806, 156

\bibitem[Tombesi(2016)]{} Tombesi, F. 2016, AN 337, 410

\bibitem[Vogelsberger(2013)]{} Vogelsberger, M., Genel, S., Sijacki, D., et al. 2013, \mnras, 436, 3013

\bibitem[Yuan(2014)]{} Yuan, F., Narayan, R. 2014, ARA\&A 52, 529

\bibitem[Zakamska(2016)]{} Zakamska, N.L., Hamann, F., P\^aris, I., et 
  al. 2016, \mnras, 459, 3144


\end{thebibliography}
\end{document}